\documentclass[twocolumn,prb]{revtex4}
\usepackage{amsfonts}
\usepackage[T1]{fontenc}
\usepackage{amsmath,amsbsy,amssymb,graphicx}
\usepackage{times}

\begin{document}

\title{ Exact solutions in two-dimensional topological superconductors:\\
Hubbard interaction induced spontaneous symmetry breaking}
\author{Motohiko Ezawa}
\affiliation{Department of Applied Physics, University of Tokyo, Hongo 7-3-1, 113-8656,
Japan}

\begin{abstract}
We present an exactly solvable model of a spin-triplet $f$-wave topological
superconductor on the honeycomb lattice in the presence of the Hubbard
interaction for arbitrary interaction strength. 
First we show that the Kane-Mele model with the corresponding spin-triplet $f$-wave superconducting pairings becomes a full-gap topological
superconductor possessing the time-reversal symmetry. We then introduce the
Hubbard interaction. The exactly solvable condition is found to be the
emergence of perfect flat bands at zero energy. They generate infinitely many conserved
quantities. It is intriguing that the Hubbard interaction breaks the
time-reversal symmetry spontaneously. As a result, the system turns into a
trivial superconductor. We demonstrate this topological property based on
the topological number and by analyzing the edge state in nanoribbon
geometry.
\end{abstract}

\maketitle

\textit{Introduction:} Topological superconductors have been investigated
intensively in this decade\cite{Qi,Sato}. A particular feature is that they
host Majorana fermions\cite{Alicea,Been,Elli}. It is a crucial problem how
the topological properties are affected by the presence of the interaction.
It is in general a formidable task to attack this problem in strongly
correlated systems. Nevertheless, if there are exactly solvable models, they
are quite powerful since they provide us with a clear physical
understanding. Exact solutions in one-dimensional Kitaev topological
superconductors have been constructed\cite{Katsura,FuChun,Ezawa,McG,Wang}.
On the other hand, as far as we are aware of, there are so far no exact
solutions for higher dimesional topological superconductors.

A Kitaev spin liquid\cite{Kitaev} on the honeycomb lattice is a beautiful
example of the exact solvable model on interacting Majoroana fermions. A key
point is that two Majorana fermion operators are made C numbers on the basis
of infinitely many conserved quantities present. Then the interacting
Majorana fermion model is transformed into a free Majorana fermion model. It
is recently shown that this method is also applicable to trivial BCS
superconductors with the Hubbard interaction\cite{Ng}. Explicit examples
have been constructed for the square and cubic lattices.

In this paper, we investigate interacting two-dimensional topological
superconductors on the honeycomb lattice. Our observation is that the
exactly solvable condition is the emergence of perfect flat bands at zero
energy. The operator degrees of freedom associated with the 
perfect flat bands become C-numbers, and generate infinitely many conserved
quantities. The resultant Majorana fermion model is transformed into a
free Majorana fermion model. 

\begin{figure}[t]
\centerline{\includegraphics[width=0.48\textwidth]{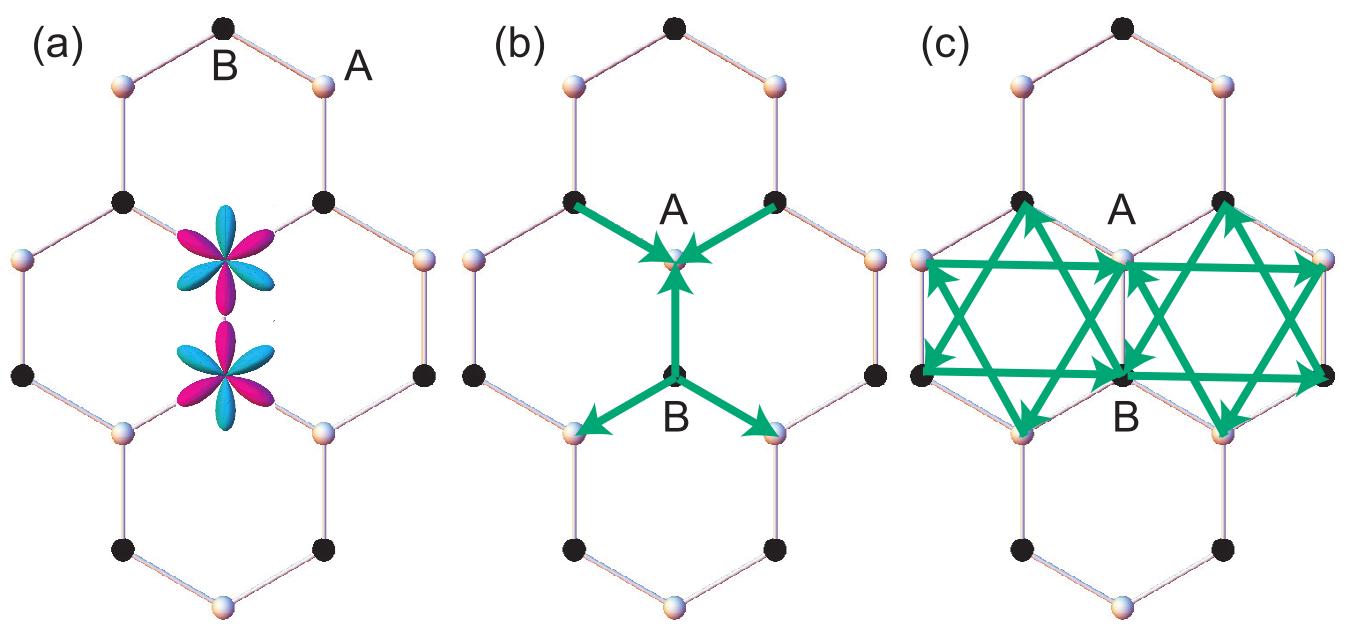}}
\caption{(a) Illustration of the honeycomb lattice and the $f$-wave
superconducting pairing. We see that the $f$-wave superconducting pairing is
compatible with the honeycomb lattice. Superconducting pairings indicated by
green arrows between (b) the nearest-neighbor sites and (c) the next-nearest
neighbor sites. The sign is plus (minus) for the forward (backward)
direction of the arrows. Arrows are pointing inside for A sites, while they
are pointing outside for B sites in (b).}
\label{FigIllust}
\end{figure}

We explicitly analyze a Kane-Mele model with the spin-triplet $f$-wave
superconductor, since the $f$-wave superconducting pairing is compatible
with the honeycomb lattice structure [Fig.\ref{FigIllust}]. We can tune this
model so as to possess perfect flat bands at zero energy.
The system becomes a full-gap superconductor.   It is shown to be a time-reversal invariant topological
superconductor by analytically evaluating the spin-Chern number. Next, we
introduce the Hubbard interaction into this system. The system is still
exactly solvable for arbitrary interaction strength. It is intriguing that
the time-reversal symmetry is spontaneously broken by the choice of the
ground state. This leads to the conclusion that the system becomes a trivial
superconductor, which we confirm by analyzing the edge states of zigzag
nanoribbons.

\textit{Model:} We start with the Kane-Mele model\cite{Kane} $H_{\text{KM}}$
on the honeycomb lattice with the spin-triplet $f$-wave superconducting
pairing. As shown in Fig.\ref{FigIllust}, the $f$-wave superconducting
pairing is compatible with the honeycomb lattice and is natural as well as
the $s$-wave superconducting pairing. We assume an equal spin pairing. The
spin-triplet $f$-wave superconductors are realized by introducing the
following interaction term\cite{Xiao} $H_{\text{SC}}$. We then introduce the
Hubbard interaction $H_{\text{Hubbard}}$. The total Hamiltonian is given by 
\begin{equation}
H=H_{\text{KM}}+H_{\text{SC}}+H_{\text{Hubbard}}  \label{HamilA}
\end{equation}%
with%
\begin{align}
H_{\text{KM}}=& -t\sum_{\left\langle i,j\right\rangle s}c_{is}^{\dagger
}c_{js}+i\frac{\lambda }{3\sqrt{3}}\sum_{\left\langle \!\left\langle
i,j\right\rangle \!\right\rangle s}\nu _{ij}\sigma _{z}c_{is}^{\dagger
}c_{js},  \label{HKM} \\
H_{\text{SC}}=& -\Delta _{1}\sum_{\left\langle i,j\right\rangle
s}c_{is}^{\dagger }c_{js}^{\dagger }+i\frac{\Delta _{2}}{3\sqrt{3}}%
\sum_{\left\langle \!\left\langle i,j\right\rangle \!\right\rangle s}\nu
_{ij}\sigma _{z}c_{is}^{\dagger }c_{js}^{\dagger }  \notag \\
& +\text{h.c.,}  \label{HSC} \\
H_{\text{Hubbard}}& =U\sum_{i}\left( c_{i\uparrow }^{\dagger }c_{i\uparrow }-%
\frac{1}{2}\right) \left( c_{i\downarrow }^{\dagger }c_{i\downarrow }-\frac{1%
}{2}\right) ,
\end{align}%
where $c_{is}^{\dagger }$ creates an electron with spin polarization $s$ at
site $i$, and $\left\langle i,j\right\rangle /\left\langle \!\left\langle
i,j\right\rangle \!\right\rangle $ run over all the nearest/next-nearest
neighbor hopping sites.\textbf{\ }We adopt the convention that $s=\uparrow
,\downarrow $ in indices and $s=+1,-1$ in equations for the up and down
spins. We explain each term: (i) The first term of the Hamiltonian $H_{\text{%
KM}}$ represents the usual nearest-neighbor hopping with the transfer
energy. (ii) The second term of the Hamiltonian $H_{\text{KM}}$ represents
the Kane-Mele spin-orbit coupling, where $\mathbf{\sigma }=(\sigma
_{x},\sigma _{y},\sigma _{z})$ is the Pauli matrix of spin, with $\nu
_{ij}=+1$ if the next-nearest-neighboring hopping is anticlockwise and $\nu
_{ij}=-1$ if it is clockwise with respect to the positive $z$ axis. (iii)
The first/second term of the Hamiltonian $H_{\text{SC}}$ represents the
nearest/next-nearest neighbor $\mathit{f}$-wave superconducting pairings.
The system has the time-reversal symmetry.

\begin{figure}[t]
\centerline{\includegraphics[width=0.48\textwidth]{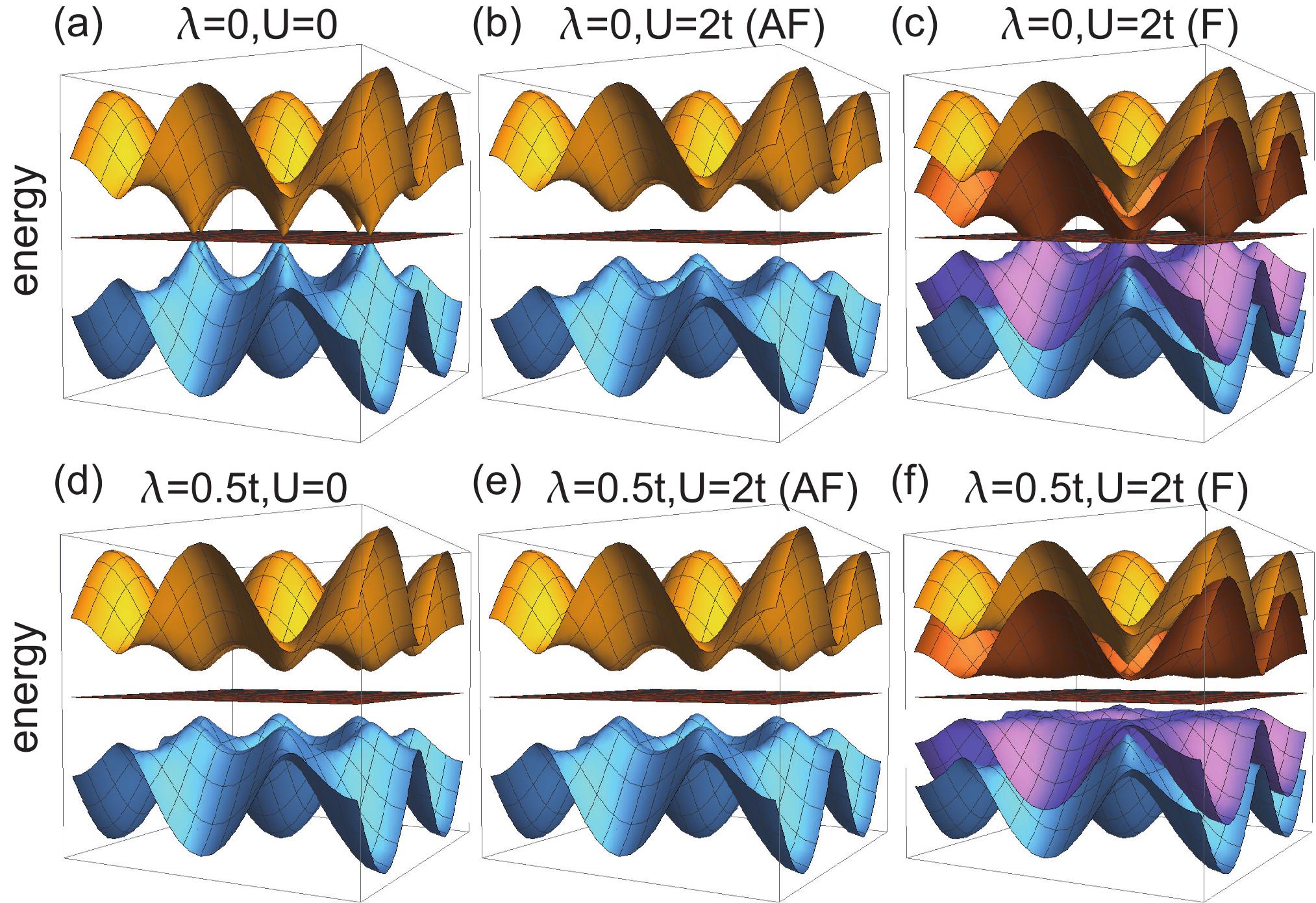}}
\caption{Bird's eye's views of the bulk band structures on the $k_x$-$k_y$
plane, where (F) indicates the ferro-order, while (AF) indicates the
antiferro-order. The energy of the antiferro-order is lower than that of the
ferro-order. The vertical axis is the energy.}
\label{FigBulk}
\end{figure}

\textit{Non-interacting case:} First we investigate the non-interacting
case; $U=0$. The Hamiltonian is block diagonal with respect to the spin; $%
H=H_{\uparrow }+H_{\downarrow }$. In order to study superconductivity, we
use the Bogoliubov de Gennes formalism. The honeycomb lattice is bipartite,
which consists of the $A$\ and $B$\ sublattices. The nearest neighbor
superconducting pairing occurs between the $A$\ and $B$\ sublattices, while
the next-nearest neighbor superconducting pairing occurs within one
sublattice: See Figs.\ref{FigIllust}(b)--(c). The Nambu spinor consists of
the electrons and holes, and reads $\Psi _{s}=\left( c_{As}\left( k\right)
,c_{Bs}\left( k\right) ,c_{As}^{\dagger }\left( -k\right) ,c_{Bs}^{\dagger
}\left( -k\right) \right) $. In the Nambu spinor basis the Hamiltonian $%
H_{s} $ has the form%
\begin{equation}
H_{s}=\left( 
\begin{array}{cccc}
s\lambda S & tF & -s\Delta _{2}S & \Delta _{1}F \\ 
tF^{\ast } & -s\lambda S & -\Delta _{1}F^{\ast } & -s\Delta _{2}S \\ 
-s\Delta _{2}S & -\Delta _{1}F & s\lambda S & tF \\ 
\Delta _{1}F^{\ast } & -s\Delta _{2}S & -tF^{\ast } & -s\lambda S%
\end{array}%
\right) ,
\end{equation}%
with 
\begin{align}
F& =e^{-iak_{y}/\sqrt{3}}+2e^{iak_{y}/2\sqrt{3}}\cos \frac{ak_{x}}{2}, \\
S& =\frac{2\lambda }{3\sqrt{3}}\left[ \sin ak_{x}-\sin \left( \frac{ak_{x}}{2%
}+\frac{\sqrt{3}ak_{y}}{2}\right) \right.  \notag \\
& \left. -\sin \left( \frac{ak_{x}}{2}-\frac{\sqrt{3}ak_{y}}{2}\right) %
\right] .
\end{align}%
The eigenvalues read%
\begin{equation}
E=\pm \sqrt{\left( t-\Delta _{1}\right) ^{2}\left\vert F\right\vert
^{2}+\left( \lambda -\Delta _{2}\right) ^{2}S^{2}}.
\end{equation}%
It is remarkable that perfect flat bands emerge when%
\begin{equation}
\Delta _{1}=t,\qquad \Delta _{2}=\lambda .  \label{ExactCondi}
\end{equation}%
As we verify later, the emergence of perfect flat bands generates infinitely
many conserved quantities and make one species of the Majorana fermions
inactive in the theory with the interaction ($U\not=0$): See (\ref{Hb}). 
In the following, we investigate the system by
requiring this condition. When $\lambda =0$, the gap closes linearly at the $K$ and $K^{\prime }$ points. 
It is a Dirac nodal superconductor: See Fig.\ref{FigBulk}(a). 
Once the $\lambda $ becomes non-zero, the system becomes a full-gap superconductor: See Fig.\ref{FigBulk}(d).

\textit{Topological number:} We show that it is a topological superconductor
by evaluating the topological numbers. They are the Chern number $\mathcal{C}
$ and the spin-Chern number $\mathcal{C}_{\sigma }$. The Chern number is
zero, $\mathcal{C}=0$, due to the time-reversal symmetry. The spin-Chern
number $\mathcal{C}_{\sigma }$ is given by $\mathcal{C}_{\sigma }=\left( 
\mathcal{C}_{\uparrow }-\mathcal{C}_{\downarrow }\right) /2$ in terms of the
spin-dependent Chern number $\mathcal{C}_{s}$ for each spin subsector\cite%
{Sheng1,Sheng2,Yan,JPSJ}. We find that $\mathcal{C}_{\sigma }$ is zero for
the flat bands since the eigen functions are given by $\Psi =\left(
1,0,1,0\right) $ and $\Psi =\left( 0,-1,0,1\right) $ for them. In order to
evaluate $\mathcal{C}_{s}$ for the valence band, we make a Taylor expansion%
\begin{equation}
F=\hbar v_{\mathrm{F}}\left( \xi k_{x}-ik_{y}\right) ,\quad S=\xi \lambda ,
\end{equation}%
where $v_{\text{F}}=\frac{\sqrt{3}}{2\hbar }at$ is the Fermi velocity and $%
\xi =1$ for the $K$ point and $\xi =-1$ for the $K^{\prime }$ point. The
Berry curvature is calculated as 
\begin{equation}
\Omega _{s}\left( \mathbf{k}\right) =\nabla \times \mathbf{a}\left( \mathbf{k%
}\right) =\frac{s\lambda }{2\left( \left( \hbar v_{\text{F}}k\right)
^{2}+\lambda ^{2}\right) ^{3/2}},
\end{equation}%
with the Berry connection $a_{i}\left( \mathbf{k}\right) =-i\left\langle
\psi \left( \mathbf{k}\right) \right\vert \frac{\partial }{\partial k_{i}}%
\left\vert \psi \left( \mathbf{k}\right) \right\rangle $. The $\mathcal{C}%
_{s}$ is explicitly evaluated as 
\begin{equation}
\mathcal{C}_{s}=\int \Omega _{s}\left( \mathbf{k}\right) d^{2}\mathbf{k/}%
\left( 2\pi \right) =\text{sgn}\left( s\lambda \right) /2,
\end{equation}%
and hence we obtain 
\begin{equation}
\mathcal{C}_{\sigma }=\text{sgn}\left( \lambda \right) ,
\end{equation}%
by adding the contributions from the $K$ and $K^{\prime }$ points.
Consequently, the system is topological for $\lambda \neq 0$. We note that
the $Z_{2}$ index is identical to the spin-Chern number provided that the
time-reversal symmetry is present and that the spin is a good quantum
number. We may explicitly confirm the Chern number, $\mathcal{C}=\mathcal{C}%
_{\uparrow }+\mathcal{C}_{\downarrow }=0$, as required by the time-reversal
symmetry.

\textit{Interacting case:} It is in general a very hard task to solve the problem
involving the Hubbard interaction $H_{\text{Hubbard}}$, since it
is not a quadratic interaction. However, as we soon see, the Hubbard
interaction can be rewritten in the quadratic form under the flat band
condition (\ref{ExactCondi}).

We introduce Majorana fermion operators $\eta$ and $\gamma$ for each sublattice defined by%
\begin{align}
c_{is}& =\eta _{is}+i\gamma _{is},\qquad c_{is}^{\dagger }=\eta
_{is}-i\gamma _{is}, \\
c_{js}& =\gamma _{js}+i\eta _{js},\qquad c_{js}^{\dagger }=\gamma
_{js}-i\eta _{js},
\end{align}%
for fermions on $i\in A$ and $j\in B$ sublattices. The Hamiltonian (\ref%
{HamilA}) is rewritten in terms of these Majorana fermions as 
\begin{equation}
H_{\text{M}}=H^{\left( 1\right) }+H^{\left( 2\right) }+H_{\text{Hubbard}}
\label{HamilB}
\end{equation}%
with%
\begin{align}
H^{\left( 1\right) }& =2i\sum_{\left\langle i,j\right\rangle s}[(\Delta
_{1}+t)\gamma _{is}\gamma _{js}+(\Delta _{1}-t)\eta _{is}\eta _{js}],
\label{H1} \\
H^{\left( 2\right) }& =\frac{1}{6\sqrt{3}}\sum_{\left\langle \!\left\langle
i,j\right\rangle \!\right\rangle s}\nu _{ij}\sigma _{z}[(\Delta _{2}+\lambda
)\gamma _{is}\gamma _{js}  \notag \\
& \quad \quad \quad \quad \quad \quad \quad \quad +(\Delta _{2}-\lambda
)\eta _{is}\eta _{js}],  \label{H2} \\
H_{\text{Hubbard}}& =U\sum_{i}\left( 2i\eta _{i\uparrow }\gamma _{i\uparrow
}\right) \left( 2i\eta _{i\downarrow }\gamma _{i\downarrow }\right) .
\label{HH}
\end{align}%
The time-reversal symmetry $\mathcal{T}$ acts as\cite{TRS}%
\begin{equation}
\mathcal{T}^{-1}i\eta _{i\uparrow }\eta _{i\downarrow }\mathcal{T}=-i\eta
_{i\uparrow }\eta _{i\downarrow },\quad \mathcal{T}^{-1}i\gamma _{i\uparrow
}\gamma _{i\downarrow }\mathcal{T}=-i\gamma _{i\uparrow }\gamma
_{i\downarrow }.
\end{equation}%
It is remarkable that the $\eta $ Majorana fermions are decoupled from the
Hamiltonians (\ref{H1}) and (\ref{H2}) by requiring the flat band condition (%
\ref{ExactCondi}). Furthermore, when we define $D_{i}=2i\eta _{i\uparrow
}\eta _{i\downarrow }$, $D_{i}$ commutes with the Hubbard interaction (\ref%
{HH}). Hence we obtain $\left[ D_{i},H_{\text{M}}\right] =0$ for all sites $i
$, which implies that $D_{i}$ becomes a C number. It is given by $D_{i}=\pm
1/2$ by using the relation $D_{i}^{2}=1/4$. It is analogous to the honeycomb
Kitaev spin liquid model\cite{Kitaev}. Substituting $D_{i}$ to (\ref{HH}),
the Hubbard interaction term becomes quadratic in terms of the Majorana
fermion operators and we obtain 
\begin{align}
H& =4it\sum_{\left\langle i,j\right\rangle s}\gamma _{is}\gamma _{js}+\frac{%
i\lambda }{3\sqrt{3}}\sum_{\left\langle \!\left\langle i,j\right\rangle
\!\right\rangle s}\nu _{ij}\sigma _{z}\gamma _{is}\gamma _{js}  \notag \\
& -U\sum_{i}D_{i}\left( 2i\gamma _{i\uparrow }\gamma _{i\downarrow }\right) .
\label{Hb}
\end{align}%
This Hamiltonian is exactly solvable for any fixed set of $D_{i}$, since it
is quadratic in terms of the fermion operators $\gamma$. A set of $D_{i}$ serves as
infinitely many conserved quantities that make the system exactly solvable.

There are two possibilities in the choice of the ground state with respect
to the $\eta $ Majorana fermions. One is the ferro-order where the sign of $%
D_{i}$ is the same between the two sublattices $D_{i}=1/2$, while the other
is the antiferro-order where the sign of $D_{i}$ is opposite between the two
sublattices $D_{i}=(-1)^{i}/2$. The $D_{i}=1/2$ and $D_{i}=-1/2$ are related
by the interchange of $U$ and $-U$. Consequently, the time-reversal symmetry
is spontaneously broken by a particular choice of $D_{i}$\ for the ground
state.

In the basis $\Psi =(\Psi _{\uparrow },\Psi _{\downarrow })$, the
Hamiltonian is written in the form%
\begin{equation}
H=\left( 
\begin{array}{cc}
H_{\uparrow } & H_{U} \\ 
H_{U}^{\ast } & H_{\downarrow }%
\end{array}%
\right) ,
\end{equation}%
with%
\begin{equation}
H_{U}=\frac{iU}{4}\left( 
\begin{array}{cccc}
1 & 0 & -1 & 0 \\ 
0 & \pm 1 & 0 & \pm 1 \\ 
-1 & 0 & 1 & 0 \\ 
0 & \pm 1 & 0 & \pm 1%
\end{array}%
\right) ,
\end{equation}%
where the sign is $+$ for the ferro-order and $-$ for the antiferro-order.
For the ferro-order, the eigenenergy is given by%
\begin{equation}
E^{2}=\left( t\left\vert F\right\vert \pm U/4\right) ^{2}+\lambda ^{2}S^{2},
\end{equation}%
and for the antiferro-order, it is given by%
\begin{equation}
E^{2}=t^{2}\left\vert F\right\vert ^{2}+\lambda ^{2}S^{2}+U^{2}/16,
\end{equation}%
where the spectrum is doubly degenerated.

By using the relation,%
\begin{align}
& 2\sqrt{t^{2}\left\vert F\right\vert ^{2}+\lambda ^{2}S^{2}+U^{2}/16}%
-\sum_{\pm }\sqrt{\left( t\left\vert F\right\vert \pm U/4\right)
^{2}+\lambda ^{2}S^{2}}  \notag \\
& =\frac{t^{2}\left\vert F\right\vert ^{2}}{\left( t^{2}\left\vert
F\right\vert ^{2}+\lambda ^{2}S^{2}\right) ^{2/3}}\frac{U^{2}}{16}+O\left(
U^{4}\right) ,
\end{align}%
we find that the antiferro-order is the ground state in the second order of $U$. 
Actually this is the case for arbitrary $U$ by numerically evaluating
the energies and comparing them. The gap closes for the ferro-order with $\lambda =0$. 
The system is a loop-nodal superconductor: See Fig.\ref{FigBulk}(c). 
On the other hand, the gap does not close for the ferro-order with $\lambda \neq 0$ [see Fig.\ref{FigBulk}(b)] 
and the antiferro-order with arbitrary $\lambda $ [see Fig.\ref{FigBulk}(e) and (f)]. The band gap is
given by $\sqrt{\lambda ^{2}+U^{2}/16}$ for the antiferro-order. In addition
to the above bands, there are perfect flat bands at zero energy, which do
not change at all by the inclusion of the Hubbard interaction. This is
because the flat bands are solely associated with the $\eta $ Majorana fermions,
which are irrelevant to the Hubbard interaction consisting solely of the $%
\gamma $ Majorana fermions.

\begin{figure}[t]
\centerline{\includegraphics[width=0.48\textwidth]{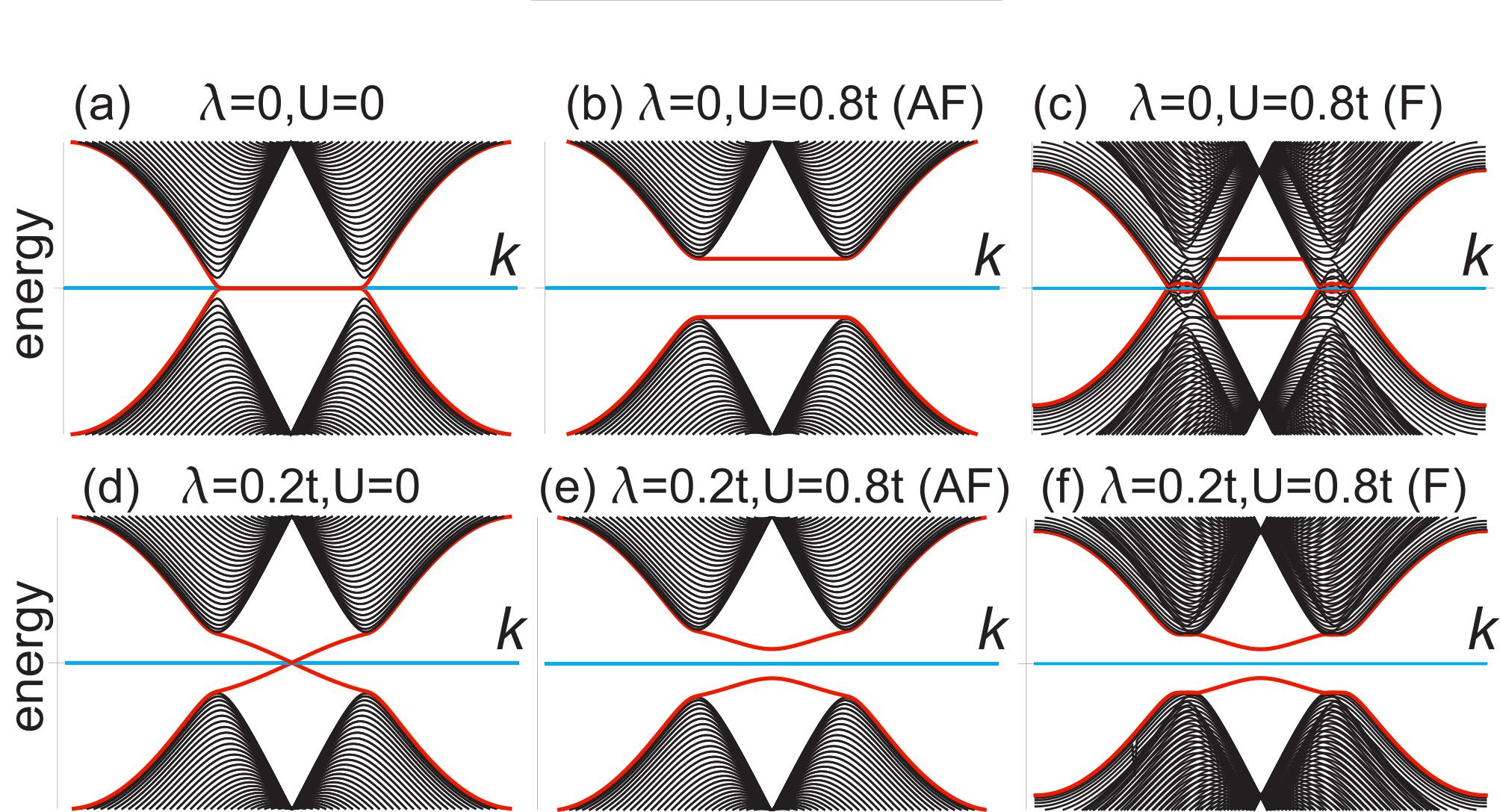}}
\caption{The band structures of zigzag nanoribbons. The horizontal axis is
the momentum $k$ and the vertical axis is the energy. Perfect flat bands at
zero energy are indicated by cyan, while the edge states are marked in
magenta.}
\label{FigRibbon}
\end{figure}

The spin-Chern number $\mathcal{C}_{\sigma }$ is no longer a good
topological number once the Hubbard interaction is switched on and a
particular choice of the ground state is made. Indeed, the time-reversal
symmetry is broken due to the last term of Eq.(\ref{Hb}) for a fixed
value of $D_{i}$ for $i$. The $Z_{2}$ index is also ill defined since the
time-reversal symmetry is spontaneously broken. On the other hand, the Chern
number is well defined and remains to be zero ($\mathcal{C}=0$) even if the
spin is no longer a good quantum number. We can check that the Chern numbers
of the two valence bands are precisely cancelled. We expect that there are
no other topological numbers in the present system. Then, the system turns
into a trivial superconductor without gap closing due to the symmetry
breaking\cite{Gap}. It is possible to confirm this by calculating the edge states in
nanoribbon geometry based on the bulk-edge correspondence.

\textit{Edge states:} In order to verify whether the system is topological
or not, we calculate the band structure for nanoribbon geometry. The band
structure of nanoribbons is given in Fig.\ref{FigRibbon}. The band
structures resemble those of graphene\cite{Graphene} for $\lambda =0$ [see
Fig.\ref{FigRibbon}(a)] and silicene\cite{Silicene} for $\lambda \neq 0$
[see Fig.\ref{FigRibbon}(d)] with the sole difference being the emergence of
perfect flat bands at zero energy.

First, we investigate the case with $\lambda =0$. There are partial flat
bands connecting the $K$ and $K^{\prime }$ points in addition to the perfect
flat bands. These partial flat bands move away from zero energy once the
interaction is introduced, as shown in Figs.\ref{FigRibbon}(b) and (c). The
bulk spectrum is fully gapped for the antiferro-order [Fig.\ref{FigRibbon}%
(b)] and gapless for the ferro-order [Fig.\ref{FigRibbon}(c)] apart from the
perfect flat bands.

Next, we investigate the case with $\lambda \neq 0$. When $U=0$, there are
helical edge states, as shown in Fig.\ref{FigRibbon}(d). These helical edge
states anticross once the interaction is introduced for both the ferro-order
[see Fig.\ref{FigRibbon}(f)] and the antiferro-order [see Fig.\ref{FigRibbon}%
(e)], which implies that the system turns into a trivial superconductor.
This can be understood in the low-energy continuum theory. The helical edge
states are perfectly localized in one of the sublattices. The low-energy
Hamiltonian is effectively given by 
\begin{align}
H_{\text{edge}}& =ivk[\gamma _{\uparrow }(k)\gamma _{\uparrow }(-k)-\gamma
_{\downarrow }(k)\gamma _{\downarrow }(-k)]  \notag \\
& -iUD_{i}\gamma _{\uparrow }(k)\gamma _{\downarrow }(-k),
\end{align}%
where $v$ is the velocity of the helical edge. The eigenenergy reads 
\begin{equation}
E=\pm \sqrt{v^{2}k^{2}+U^{2}},
\end{equation}%
which well describes the anticrossing of the helical edge states: See Figs.\ref{FigRibbon}(e) and (f). 
We note that the anticrossed edge states are
almost the same between the ferro-order and the antiferro-order although the
bulk spectrum is different, which implies the above low-energy theory is
valid.

We have constructed exact solvable models on two-dimensional topological
superconductors. The exact solvable condition is the emergence of perfect
flat bands at zero energy, generating infinitely many conserved quantities
and making one species of the Majorana fermions inactive. This method will
be a guideline for searching further exact solvable models.

The author is very much grateful to N. Nagaosa for many helpful discussions
on the subject. This work is supported by the Grants-in-Aid for Scientific
Research from MEXT KAKENHI (Grant Nos.JP17K05490 and 15H05854). This work is
also supported by CREST, JST (JPMJCR16F1).

\end{document}